\documentclass[aps,pra,showpacs, preprint, superscriptaddress, amsmath, amssymb, floatfix]{revtex4-1}
\usepackage{graphicx}
\usepackage{epstopdf}
\usepackage{subfigure}
\usepackage{dcolumn}
\usepackage{bm}
\usepackage{extarrows} 
\usepackage{hyperref}
\hypersetup{colorlinks=true,       
   linkcolor=blue,          
   citecolor=blue,        
   urlcolor=blue }	

\def\be{\begin{equation}}      
\def\ee{\end{equation}}
\def\bea{\begin{eqnarray}}      
\def\eea{\end{eqnarray}}

\def\beu{\begin{equation*}}   
\def\eeu{\end{equation*}}

\providecommand{\mr}[1]{\mathrm{#1}}
\providecommand{\abs}[1]{\left\lvert#1\right\rvert}   

\providecommand{\pp}[2]{\frac{\partial{#1}}{\partial{#2}}}

\providecommand{\fr}[2]{\frac{#1}{#2}}

\usepackage[usenames,dvipsnames]{color}    
\usepackage{soul}   
\setstcolor{red} 
\setul{}{1.5pt}  

\begin{document}
\title{A quasi-mode theory of chiral phonons}


\author{Xunnong Xu }
\affiliation{Joint Quantum Institute, University of Maryland/National Institute of Standards and Technology, College Park, Maryland 20742, USA}
\author{Seunghwi Kim}
\affiliation{Mechanical Science and Engineering, University of Illinois at Urbana-Champaign Urbana, Illinois 61801, USA} 
\author{Gaurav Bahl}
\affiliation{Mechanical Science and Engineering, University of Illinois at Urbana-Champaign Urbana, Illinois 61801, USA} 
\author{Jacob M. Taylor}
\affiliation{Joint Quantum Institute, University of Maryland/National Institute of Standards and Technology, College Park, Maryland 20742, USA}
\affiliation{Joint Center for Quantum Information and Computer Science, University of Maryland, College Park, Maryland 20742, USA}
\affiliation{Research Center for Advanced Science and Technology (RCAST), The University of Tokyo, Meguro-ku, Tokyo, 153-8904, Japan}



\date{\today}
\begin{abstract}
The coherence properties of mechanical resonators are often limited by multiple unavoidable forms of  loss –- including phonon-phonon and phonon-defect scattering -- which result in the scattering of sound into other resonant modes and into the phonon bath. Dynamic suppression of this scattering loss can lift constraints on device structure and can improve tolerance to defects in the material, even after fabrication. 
Inspired by recent experiments, here we introduce a model of phonon losses resulting from disorder in a whispering gallery mode resonator with acousto-optical coupling between optical and mechanical modes.
We show that a typical elastic scattering mechanism of high quality factor (Q) mechanical modes flips the direction of phonon propagation via high-angle scattering, leading to damping into modes with the opposite parity. 
When the optical mode overlaps co-propagating high-Q and bulk mechanical modes, the addition of laser cooling via sideband-resolved damping of the mechanical mode of a chosen parity  also damps and modifies the response of the bulk modes of the same parity. This, in turn, simultaneously improves the quality factor and reduces the thermal load of the counter-propagating high-Q modes, leading to the dynamical creation of a cold phononic shield. We compare our theoretical results to the recent experiments of Kim et al., and find quantitative agreement with our theory.

\end{abstract}


\maketitle

\section{Introduction} 

Quantum optomechanics studies the radiation pressure-mediated interaction between light and mechanical motion in the quantum regime \cite{Caves1980, Braginsky2001, Braginsky2002, Kippenberg2005}. Crucial experimental advances \cite{Kippenberg2008, Marquardt2009, Aspelmeyer2014} make investigation of non-classical effects possible, as evidenced by recent results such as the ground state cooling of mechanical resonator \cite{WilsonRae2007, Teufel2011, Chan2011}, the generation of  squeezed states of light \cite{Brooks2012, SafaviNaeini2013, Purdy2013}, and studies of single-photon nonlinear optics \cite{Nunnenkamp11, Rabl2011, Kronwald2013, Lemonde2013, Borkje2013, Xu2015} in various optomechanics platforms. 
An example optomechanical interaction occurs in Brillouin scattering (BS), where acoustic vibrations are induced by acoustic-optic coupling \cite{Grudinin2009, Grudinin2010, Bahl2013}, and optomechanical cooling via  Brillouin scattering  has already been demonstrated \cite{Bahl2012}. 

In a recent experiment by Kim {\it et.al.} \cite{Kim2016nature}, the chiral behavior of phonons in silica microsphere resonator was observed.  We showed that the phonon mode co-propagating with the pump laser is optomechanically cooled via forward Brillouin scattering and its linewidth has been broadened, while the linewidth of the counter-propagating phonon mode is made narrower. This signature of chirality -- broken symmetry of the scattering properties of co-propagating/counter-propagating phonons -- becomes more evident when the input optical power of the pump laser is increased. As an analogy to emerging studies of chiral photonics interacting with emitters \cite{Pichler15, Ramos2016, Vermersch2016}, the observed chiral phonon broken symmetry in a microsphere resonator could be a candidate platform for the study of chiral networks of phononic modes. 

Here we introduce a model  to describe these experiments. We primarily consider the high-angle elastic scattering between a particular high quality factor (Q) mechanical mode of interest and a continuum of lossy bulk mechanical modes.  We find that when the optical coupling cools the co-propagating high-Q and bulk mechanical modes, the optical modification of scattering between these bulk modes and the counter-propagating high-Q mode simultaneously improves the quality factor and reduces the thermal load of the same counter-propagating high-Q mode, leading to the dynamical creation of a cold `phononic shield' \cite{Chan2012}.

We detail our theoretical model  in Section II. We then calculate the linewidth and effective temperature of phonons using linear response theory in Section III to explain the key findings of the experiment \cite{Kim2016nature}. Finally, we fit the linewidth data of the experimental and estimate some of the key parameters in Section IV.

\section{Model}
\subsection{Photon-phonon interaction}
\begin{figure}[!h]
\begin{center}
\includegraphics[width=.98 \columnwidth]{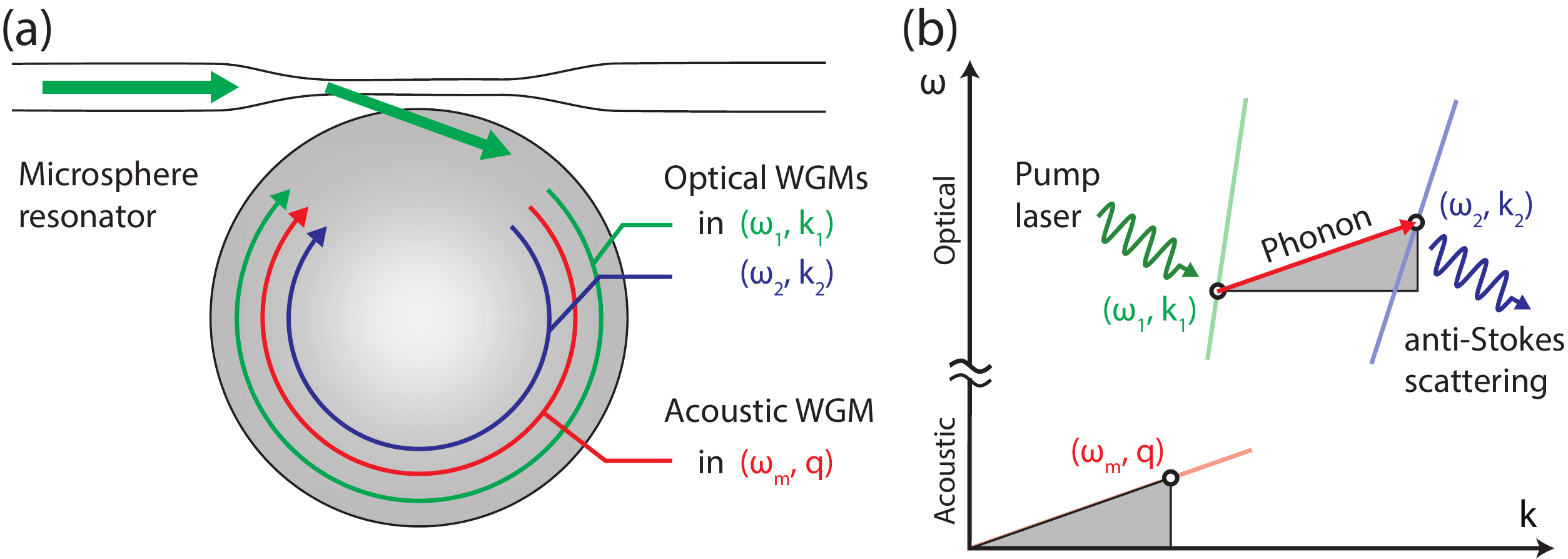}
\caption[Forward Brillouin scattering]  {(a) The generation of anti-Stokes photon (blue) and absorption of phonon (red) from a pump photon (green) via forward Brillouin scattering process. Due to symmetry, the resonator supports both degenerate pairs of  co-propagating and counter-propagating photons and phonons, but only co-propagating modes are shown. (b) Dispersion relation for phonon (solid red line) and photons (solid green, blue lines), and the mechanism of coherent acousto-optical interaction. }
\label{fig1}
\end{center}
\end{figure}

We consider a model of acousto-optic interaction in a multi-mode whispering gallery-type resonator that supports photon and phonon modes. Photons from an adjacent waveguide are interfaced with the resonator optical modes through evanescent coupling.  Phonons that occupy a surface acoustic wave resonator mode are annihilated via forward Brillouin scattering process along with the creation of anti-Stokes scattered photons as shown in Fig.~\ref{fig1}.

We can use the rotational symmetry of the system to write the displacement field $\phi$ and electromagnetic field $\psi$ as the following forms in  cylindrical coordinates: 
\bea
\phi &=& \sum_{q, ,m}  f_{q, m} (r, z) \left[ b_{q, m} e^{iq r} + b_{-q, m} e^{-iq r} + \mathrm{H. C.}\right] \\
\psi &=& \sum_{k, m}  g_{k, m} (r, z) \left[ c_{k, m} e^{ikr} + c_{-k, m} e^{-ik r} + \mathrm{H. C.}\right]
\eea
where $f$ is a mode profile function, $b$ and $c$  are the annihilation operator for phonon and photons respectively, and $q, k, m$ are the quantum numbers representing different momentum and angular momentum eigenstates. 
The interaction between  $\phi$ and  $\psi$ comes from acoustic-optical effect: a change in the susceptibility $\epsilon$ of the material because of the strain from the displacement.  It can be calculated as 
\be
V \approx \int dr dz d\theta \frac{1}{2} \pp{\epsilon}{s} s \abs{\psi}^2, 
\ee
where the strain field $s$ is related to the displacement field $\phi$ by $s =\partial_r \phi$, and we take $\partial \epsilon/\partial s$ to be a constant determined by material properties. In explicit form, the interaction is  
\begin{widetext}
\bea
V &=&  \frac{1}{2} \pp{\epsilon}{s} \int dr dz d\theta \sum_{q, m} f_{q, m} (r, z) iq \left(b_{q,m} e^{iq r} - b_{-q, m}e^{iq r}\right) \nonumber \\
&& \times \sum_{k,k^{\prime}, m, m^{\prime}} g_{k, m}  (r, z) g_{k^{\prime}, m^{\prime}}  (r, z)\frac{\hbar\sqrt{\omega_k \omega_{k^{\prime}}} }{2\epsilon ({\mathrm{Vol}}) }\left[c_{k,m} c_{k^{\prime}, m^{\prime}} e^{i(k + k^{\prime})r } +  c_{k,m} c_{k^{\prime}, m^{\prime} }^{\dag} e^{i(k - k^{\prime})r } + \mathrm{H.C.} \right], 
\eea
\end{widetext}
where $\mathrm{Vol} $ is the effective mode volume. The integral over the exponential factors gives us a mode-matching condition for different photon-phonon interaction processes, e.g, $b_{q,m} c_{k,m} c_{k^{\prime}, m^{\prime}}^{\dag}$ for creating a new photon from annihilating a phonon and a photon pair.

\subsection{Quasi-mode picture}
We consider two nearby (in frequency) optical modes of the resonator that couple to the vibrational excitations of the underlying  medium, through Brillouin acousto-optic scattering. Labelling these modes $c_P$ (for pump) and $c$ (for the higher frequency anti-Stokes probe mode), we write the optomechanical interaction
\be
V = \sum_k (b_k + b_{-k}^\dag) (\lambda_k^* c_P^\dag c + \lambda_k c^\dag c_P) + \ldots c_P^\dag c_P + \ldots c^\dag c.  \\
\ee
where $\lambda_k$ describes coupling between these modes, which will naturally account for quasi-phase matching and other constraints. The terms with elipses in front we neglect due to a lack of phase-matching.

Upon strong optical pumping of the $c_P$ mode, we can look at the fluctuation away from the classical steady state with amplitude $\alpha$, $c_P \rightarrow \alpha + c_P$,  and similarly for $b_k$. Thus, in the limit $\alpha \gg 1$, we find (with a trivial gauge transform of $c$) \cite{Kippenberg2008, Aspelmeyer2014} 
\be
V_\alpha = \sum_k \alpha \lambda_k (b_k + b_{-k}^\dag)(c + c^\dag)
\ee
where now $\lambda_k \in \mathbb{R}$. We define the pump-enhanced coupling $\Lambda_k = |\alpha \lambda_k|$.

Let us single out two high Q mechanical modes representing a time-reversed pair of interest, which has an intrinsic degeneracy for co-propagating and counter-propagating directions, relabeling them $a_+$ and $a_-$. These are also coupled to the $b_k$ modes in a quasi-mode theory of mechanical damping. We have $c$ coupled to $a_+$ but not to $a_-$, again due to phase matching (momentum conservation).  The scattering of phonons off disorder within the material can mix phonons with different momenta \cite{Datta1997}, but we neglect the same-same scattering $b_k \leftrightarrow a_+$  since it does not break chiral symmetry, and so should be included in the definition of the achiral modes. Thus only relevant term is the $b_k \leftrightarrow a_-$ scattering. We now want to understand the mediated interaction between $a_-$ and $c$ through the coupling to the bulk modes $b_k$. 

Moving to the Fourier domain, and adding a weak thermalization of $b_k$ modes with rate $\eta$, we have the Heisenberg-Langevin equations of motion in the rotating wave approximation ($\Delta < 0, |\Delta| \gg \kappa$) \cite{Weisskopf1930, Scully1997, Weiss2012}:
\begin{subequations}
\begin{align}
-i \nu c &= i \Delta c - \frac{\kappa}{2} c + \sqrt{\kappa} c_{in} - i \sum_k \Lambda_k b_k \\
-i \nu a_- &= - i \omega_m a_- - i \sum_k \mu_k b_k\\
-i \nu b_k &= -i \omega_k b_k - \frac{\eta}{2} b_k + \sqrt{\eta} b_{k,in} - i \Lambda_k c - i \mu_k a_- 
\end{align}
\end{subequations}
with $\nu$ the Fourier frequency, $\omega_m$ the mechanical frequency, $\kappa$ the optical damping, $\mu_k$ the coupling between $a_-$ and $b_k$, and $c_{in}$ and $b_{in} $ are the corresponding input fields. 

We can solve  Eq.~(7c), getting
\bea
-i \sum_k \Lambda_k b_k  &&= -i \sum_k \Lambda_k \frac{\sqrt{\eta} b_{k,in} - i \Lambda_k c - i \mu_k a_-}{i (\omega_k - \nu) + \eta/2} \nonumber \\
&&= - \alpha^2 \sum_k \frac{\lambda_k^2}{i (\omega_k - \nu) + \eta/2} c  \nonumber \\
&& ~~ - \alpha \sum_k \frac{\lambda_k \mu_k}{i (\omega_k - \nu) + \eta/2} a_- + \sqrt{\Gamma} b_{in}
\eea
where we define a new input field
\be
\sqrt{\Gamma} b_{in} \equiv -i \alpha \sum_k \lambda_k  \frac{\sqrt{\eta}}{i (\omega_k - \nu) + \eta/2} b_{k,in}
\ee

We see that equation Eq.~(7a) has a new damping term due to the real part of the sum over $k$. Converting the sum to an integral over bath bulk modes frequencies $\sum_k = \int \rho(\omega) d\omega$, we can perform the integral within the rotating wave approximation (allowing us to take the lower bound of frequencies to minus infinity) and recover
\bea
\sum_k \frac{\lambda_k^2}{i (\omega_k - \nu) + \eta/2} 
&=& \int \rho(\omega) \lambda(\omega)^2 \frac{1}{i (\omega - \nu) + \eta/2} d\omega \nonumber \\
&=&  \pi \rho(\nu) \lambda(\nu)^2 + \mathcal{P}(\ldots)
\eea
where we have the principal value part of the integral leading to a frequency shift, while the other component leads to decay of the $c$ mode. This defines $\Lambda = 2\pi \alpha^2 \rho(\nu) \lambda(\nu)^2$. We also get a damping of $a_-$, $\gamma_{a_-} = 2\pi \rho(\nu) \mu(\nu)^2$, which leads to the backscatter-induced loss of phonons.  
However, there is a cross term in the damping, 
\be
\alpha \chi \equiv \alpha 2\pi \lambda(\nu) \mu(\nu) \rho(\nu)
\ee
suggesting interference between two decay pathways. 
We can understand that the effective $c$ and $a_-$ equations of motion are generated by an effective Hamiltonian in the stochastic Schrodinger equation sense with two imaginary terms for damping:
\be
-i \frac{\kappa}{2} c^\dag c - i \frac{1}{2}\left( \sqrt{\Lambda} c + \sqrt{\gamma_{a_-}} a_-\right)^\dag \left(\sqrt{\Lambda} c + \sqrt{\gamma_{a_-}} a_- \right)
\ee
That is, damping occurs for a superposition of the $c$ and $a_-$ mode, in the corresponding Lindblad-form superoperator.

\subsection{Simpler version of the model: a quasi-mode picture of the bulk}
This relatively complicated model above can be reduced in the rotating-wave approximation (RWA), narrowband limit to a much simpler model. Specifically, let us define a new self-consistent quasi-mode $b$. We then have for its equation of motion
\be
-i \nu b = -i \nu b - \frac{\Gamma}{2} b + \sqrt{\Gamma} b_{in} - i (\alpha \tilde\lambda c + \tilde g a_-) 
\ee
where we see that the frequency dependence drops out -- due to the continuum nature of the actual $b_k$ modes. We get the same physics as the above model if we take $\eta= \frac{\pi}{\rho}$, which is also the intrinsic damping rate $\gamma_b$ of the  $b_k$ modes. The continuum model of the quasi-mode also suggests that its validity requires that its damping rate be large compared to the intrinsic damping rate for phonons: $\Gamma \gg \gamma_b$. 

Using the quasi-mode $b$, the system supports degenerate phonon and phonon modes, with $+$ stands for the co-propagating direction and  $-$ stands for the counter-propagating direction. To summarize, we have made the following assumptions based on the continuum model: 
\begin{enumerate}
	
	\item Phonon backscattering occurs between high-Q modes and their opposite direction quasi-mode ($a_{+} \longleftrightarrow b_{-}$, $a_{-}  \longleftrightarrow b_{+}$) with strength $V_0 $, and between modes of the same type ($a_{+} \longleftrightarrow a_{-}$, $b_{+}  \longleftrightarrow b_{-}$) with strength $V_1$. 
	
	\item The optical mode $c_{+(-)}$ couples to the high-Q mode $a_{+(-)}$ and the quasi-mode $b_{+(-)}$ with different weights. Specifically, $c_+$ couples to $a_+$ via direct optomechanical interaction with strength $G = \alpha \lambda$ and couples to the quasi-mode with strength $\alpha g$, while $c_-$ couples to $a_-$ with strength $ \beta \lambda$ and couples to the quasi-mode with strength $\beta g $. 
	
	\item The mechanical modes $a_{+(-)}$ and $b_{+(-)}$ have different damping rates $\gamma$ and $\Gamma$ with $\gamma \ll \Gamma$ , but the damping is symmetric between the $\pm$ modes. We also assume that $\Gamma$ is in the same order as the optical loss rate $\kappa$, both of which are much larger than $\gamma$. 
\end{enumerate}

A model with these assumptions is shown below in Fig.~\ref{fig2}, 
\begin{figure}[!h]
\begin{center}
\includegraphics[width=.8 \columnwidth]{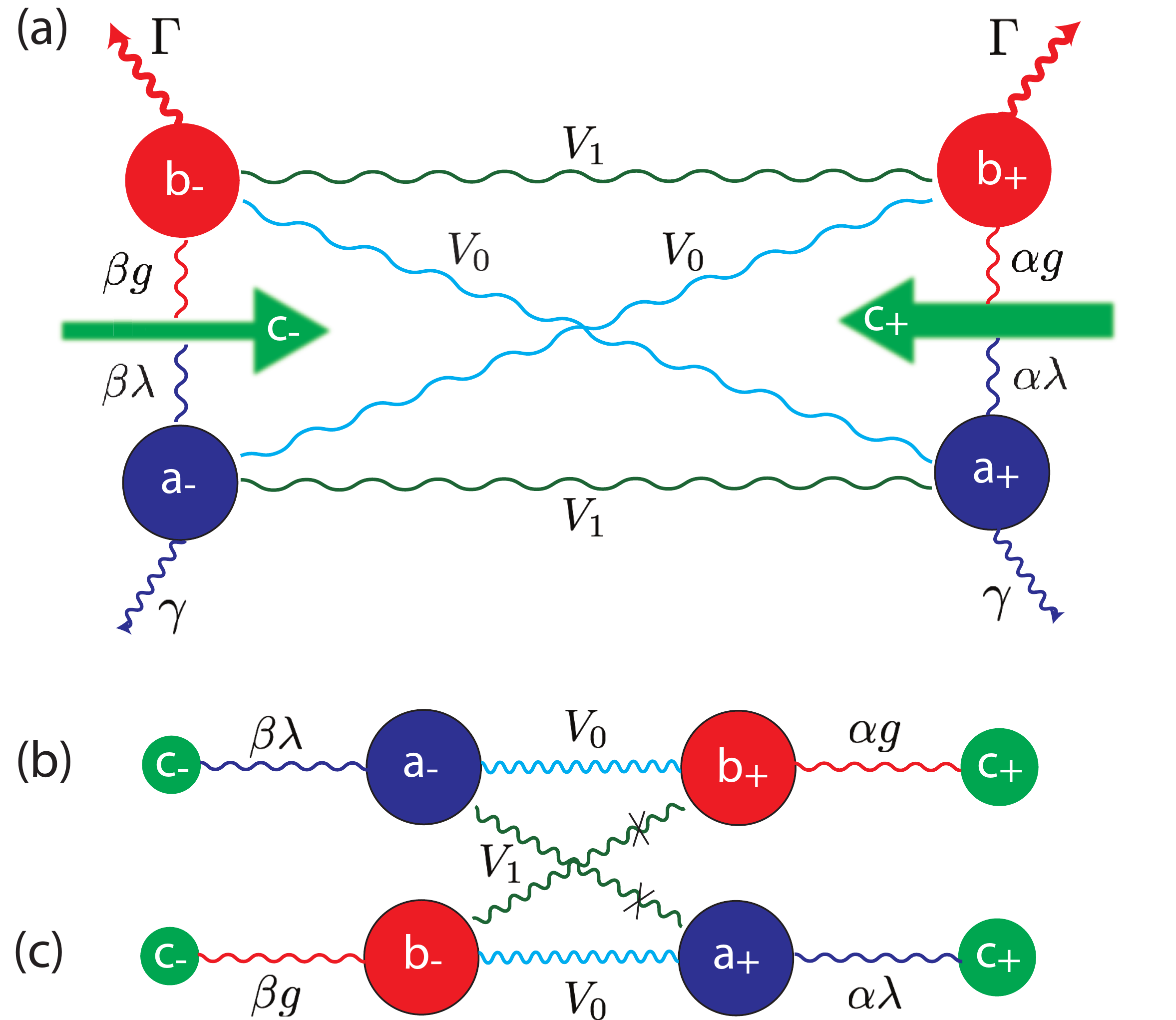}
\caption[]{Simplified multi-mode theory of optomechanical interaction and phonon scattering. (a) For each propagating direction, a quasi-mode $b$ and a particular mode $a$ are coupled to the driving field $c$ with different rates. Phonon back-scattering happens between the two directions. (b-c): We neglect the phonon scattering between the same species (the $V_1$ lines) and also assume that the optical driving fields couple to the quasi-mode and particular mode separately, since the driving field is very strong. With these assumptions, we can break the loop in $a$ into two sub diagrams. }
\label{fig2}
\end{center}
\end{figure}
and its effective hamiltonian is given by
\bea
H_{\mathrm{eff}} &=& -\Delta (c_+^{\dag}c_+ + c_-^{\dag}c_-)   + \omega_m(a_{+}^{\dag} a_{+} + a_{-}^{\dag} a_{-})   + \omega_b (b_{+}^{\dag} b_{+} + b_{-}^{\dag} b_{-}  ) + \alpha c_+^{\dag} (\lambda a_{+}  + g b_{+})  + \mathrm{h.c.}   \nonumber   \\
&&+  \beta c_-^{\dag} (\lambda  a_{-}  + g b_{-})  + \mathrm{h.c.} +  V_0 (a_{+}^{\dag} b_{-}  + a_{-}^{\dag} b_{+} + \mathrm{h.c.})  +  V_1 (a_{+}^{\dag} a_{-}  + b_{+}^{\dag} b_{-} + \mathrm{h.c.})   \nonumber\\
&& -i\frac{\kappa}{2} (c_+^{\dag}c_+ + c_-^{\dag}c_-)  -i \frac{\gamma}{2}( a_{+}^{\dag} a_{+}  + a_{-}^{\dag} a_{-})  - i\frac{\Gamma}{2}(b_{+}^{\dag} b_{+} + b_{-}^{\dag} b_{-}  ), 
\eea
where the optical and mechanical loss are modeled by anti-Hermitian hamiltonian. Again, this effective Hamiltonian is a short-hand for the full input-operator picture used in the Heisenberg-Langevin equations, below. Here we used a nominal frequency $\omega_b$ for the quasi-modes $b_{\pm}$, but when we go to the Fourier domain, its frequency dependence will drop out, in the sense that $\omega_b \to  \nu$.

While the dynamics of the system can, in principle, be solved numerically, the loop structure in this coupled six-mode system will make the result quite complicated and it is hard for us to interpret the main physics in the system without further approximation. We assume the $g$ parameter is larger than $\lambda$,  such that the optical field couples more strongly to the bulk modes $b_{\pm}$. We can then break the loop into two pieces by cutting the lines representing $V_1$ interactions, both of which consists of two mechanical modes and two optical modes, as shown in Fig.~2(b)-(c). The validity of the assumption of ignoring the $V_1$ interaction is discussed in more detail in Section III, where we show that the possible consequences of this interaction  have not been observed in experiment.

\section{System dynamics}
\subsection{Linear response} 
We now focus on Fig.~2(b) and Fig.~2(c) to calculate the linewidth for co-propagating phonon $a_+$ and counter-propagating phonon $a_-$. In the experiment \cite{Kim2016nature}, the counter-propagating pump power is about 10 times smaller than that of co-propagating pump, so we neglect the effect of counter-propagating optical mode $c_-$ first. For Fig.~2(b), as shown in our toy model, we have 
\bea
H_{\text{toy}} &=& -\Delta (c_+^{\dag}c_+ + c_-^{\dag}c_-)  + \omega_m a_{-}^{\dag} a_{-} + \omega_b b_{+}^{\dag} b_{+}  \nonumber \\
&& + \alpha g (c_+^{\dag} b_{+}  + b_{+}^{\dag} c_+) + \beta \lambda(a_- c_-^{\dagger} + c_-^{\dag} a_-)   \nonumber  \\
&& + V_0 (a_{-}^{\dag} b_{+} + b_{+}^{\dag} a_{-} )  -i\frac{\kappa}{2} (c_+^{\dag}c_+ + c_-^{\dag}c_-) \nonumber \\
&& -i \frac{\gamma}{2}a_{-}^{\dag} a_{-}  -i\frac{\Gamma}{2} b_{+}^{\dag} b_{+}. \eea
We write down the Heisenberg-Langevin equations for each mode and transform to frequency domain to solve the equations. 
\begin{subequations}
\bea
-i\nu c_+ &=& i\Delta c_+ -\frac{\kappa}{2} c_+ + \sqrt{\kappa}c_+^{\mr{in}} -i \alpha g b_+, \\
-i\nu c_- &=& i\Delta c_-  -\frac{\kappa}{2} c_- + \sqrt{\kappa}c_-^{\mr{in}} -i \beta \lambda a_-, \\
-i\nu a_-  &=& -i\omega_m a_-  - \frac{\gamma}{2} a_- + \sqrt{\gamma}a_-^{\mr{in}} - i \beta\lambda c_- - i V_0 b_+ ,  \\
-i\nu b_+ &=&  -i\nu b_+  - \frac{\Gamma}{2} b_+ + \sqrt{\Gamma} b_+^{\mr{in}} - i\alpha g c_+ - i V_0 a_-  .  
\eea
\end{subequations}
As shown before, the frequency dependence of the quasi-mode $b_{\pm}$ drops out in the Fourier domain. We now proceed to eliminate $c_+,~c_-$ and $b_+$ to understand the behavior of $a_-$. From the equation for $b_+$ Eq.~(16d) we have
\be
b_+ = \frac{\sqrt{\Gamma} b_+^{\mr{in}} - i\alpha g c_+ - i V_0 a_- }{\Gamma/2}
\ee
plug this into the equation for $c_+$,  we have an equation which only relates $c_+$ and $a_-$:
\be
\left[-i(\nu + \Delta) + \kappa/2\right]c_+ = \sqrt{\kappa} c_+^{\text{in}} - i\alpha g  \frac{\sqrt{\Gamma} b_+^{\mr{in}} - i\alpha g c_+ - i V_0 a_- }{\Gamma/2} 
\ee
This simplifies to 
\be
\left[-i(\nu + \Delta) + \kappa/2 + \frac{\alpha^2 g^2}{\Gamma/2}\right]c_+ = \sqrt{\kappa} c_+^{\text{in}} - i\frac{\alpha g}{\sqrt{\Gamma}/2}  b_+^{\mr{in}}  - \frac{ \alpha g V_0 }{\Gamma/2} a_- 
\ee
From the left hand side of the equation, we see an optomechanical modification to the damping rate for $c_+$. Also, the $c_+$ mode is effectively coupled to the $a_-$ mode via the interaction with $b_+$, which means the properties of $a_-$ could possibly be modified by a driving field in the opposite direction. We are going to analyze this in more detail in following section. We further define the susceptibility of $c_+$ as $\tilde{\kappa} \equiv -i(\nu + \Delta) + \kappa/2 + 2\alpha^2 g^2/\Gamma $, then 
\be
c_+ =  \frac{\sqrt{\kappa} c_+^{\text{in}} - i\frac{\alpha g}{\sqrt{\Gamma}/2}  b_+^{\mr{in}}  - \frac{ \alpha g V_0 }{\Gamma/2} a_- }{\tilde{\kappa}}. 
\ee
We now put this optical field back into the $b_+$ equation, we get 
\bea
b_+ &=& \frac{\sqrt{\Gamma} b_+^{\mr{in}} - i\alpha g  \cfrac{\sqrt{\kappa} c_+^{\text{in}} - i\frac{\alpha g}{\sqrt{\Gamma}/2}  b_+^{\mr{in}}  - \frac{ \alpha g V_0 }{\Gamma/2} a_- }{\tilde{\kappa}}  - i V_0 a_- }{\Gamma/2} \nonumber \\
&=& \frac{1}{\sqrt{\Gamma}/2} \left( 1 - \frac{\alpha^2 g^2}{\Gamma\tilde{\kappa} /2} \right) b_+^{\text{in}} - \frac{i\alpha g \sqrt{\kappa}}{\Gamma\tilde{\kappa}/2} c_+^{\text{in}}   - \frac{iV_0}{\Gamma/2} \left(1 - \frac{\alpha^2 g^2}{\Gamma\tilde{\kappa} /2}  \right) a_- . 
\eea
The equation for $c_-$ 
\be
c_- = \frac{\sqrt{\kappa} c_-^{\text{in}} -i\beta\lambda a_- }{-i(\nu + \Delta) + \kappa/2} ,
\ee
which indicates that $c_-$ is only modified by the coupling to $a_-$. We can put $b_+$ and $c_-$ back into the equation for $a_-$, and get
\bea
 \chi_-^{-1} a_-  &=&  \sqrt{\gamma}a_-^{\mr{in}}  -  \frac{i \beta\lambda \sqrt{\kappa} }{-i(\nu + \Delta) + \kappa/2} c_-^{\text{in}}  - \frac{V_0\alpha g \sqrt{\kappa}}{\Gamma\tilde{\kappa} /2} c_+^{\text{in}} \nonumber \\
&& - \frac{iV_0}{\sqrt{\Gamma}/2} \left( 1 - \frac{\alpha^2 g^2}{\Gamma\tilde{\kappa}/2} \right) b_+^{\text{in}} 
\eea
where 
\bea
\chi_-^{-1} &=& -i(\nu -\omega_m) + \gamma/2 + \frac{\beta^2\lambda^2}{-i(\nu + \Delta) + \kappa/2} \nonumber \\
&&+  \frac{V_0^2}{\Gamma/2} \left(1 - \frac{\alpha^2 g^2}{\Gamma\tilde{\kappa} /2}  \right)   
\eea
is the inverse susceptibility of $a_-$ to input fields. 

Similarly, for the co-propagating phonon mode $a_+$, we can find its equation of motion by interchanging $\alpha$ with $\beta$, $a_+$ with $a_-$, and $c_+$ with $c_-$: 
\bea
\chi_+^{-1} a_+ &=&  \sqrt{\gamma}a_+^{\mr{in}}  -  \frac{i \alpha\lambda \sqrt{\kappa} }{-i(\nu + \Delta) + \kappa/2} c_+^{\text{in}}  - \frac{V_0\beta g \sqrt{\kappa}}{\Gamma\tilde{\kappa}^{\prime} /2} c_-^{\text{in}} \nonumber \\
&& - \frac{iV_0}{\sqrt{\Gamma}/2} \left( 1 - \frac{\beta^2 g^2}{\Gamma\tilde{\kappa}^{\prime}/2} \right) b_-^{\text{in}} 
\eea
with $\tilde{\kappa}^{\prime} \equiv -i(\nu + \Delta) + \kappa/2 + 2\beta^2 g^2/\Gamma$ and 
\bea
 \chi_+^{-1} &=& -i(\nu -\omega_m) + \gamma/2 + \frac{\alpha^2\lambda^2}{-i(\nu + \Delta) + \kappa/2} \nonumber \\
 && +  \frac{V_0^2}{\Gamma/2} \left(1 - \frac{\beta^2 g^2}{\Gamma\tilde{\kappa}^{\prime} /2}  \right)  
\eea

\subsection{Phonon linewidth}
We can define the cooperativities as $\mathcal{C}_\alpha = 4\alpha^2 g^2/\Gamma\kappa$ and $\mathcal{C}_\beta =  4\beta^2 g^2/\Gamma\kappa$, which are both dimensionless parameters describing the strength of optomechanical coupling of the quasi-modes relative to cavity decay rate and mechanical damping rate. 

The modified linewidth of the $a_{\pm}$ phonons, as a result of optomechanical cooling/heat, is given by the real part of its susceptibility, so we have
\bea
\gamma_{a_+} (\nu) &=& 2\text{Re}(\chi_{a_+}^{-1} (\nu)  ) \nonumber \\
&=& \gamma + \frac{\alpha^2\lambda^2 \kappa}{(\nu + \Delta)^2 + \kappa^2/4}  \\
&&+ \frac{4V_0^2}{\Gamma} \left[1 - \frac{2\beta^2 g^2}{\Gamma} \frac{(1+\mathcal{C}_\beta)\kappa/2 }{(\nu + \Delta)^2 +(1+\mathcal{C}_\beta)^2 \kappa^2/4 }  \right], \nonumber 
\eea
and 
\bea
\gamma_{a_-} (\nu) &=& 2\text{Re}(\chi_{a_-}^{-1} (\nu)  ) \nonumber \\
&=& \gamma + \frac{\beta^2\lambda^2 \kappa}{(\nu + \Delta)^2 + \kappa^2/4} \\
&& +  \frac{4V_0^2}{\Gamma} \left[1 - \frac{2\alpha^2 g^2}{\Gamma} \frac{(1+\mathcal{C}_\alpha)\kappa/2 }{(\nu + \Delta)^2 +(1+\mathcal{C}_\alpha)^2 \kappa^2/4 } \right], \nonumber 
\eea
For the special case of $\nu = -\Delta$, we have
\begin{subequations}
\bea
\gamma_{a_+} &=&\gamma + \frac{4\alpha^2\lambda^2}{\kappa} +  \frac{4V_0^2}{\Gamma}\frac{\kappa}{ \kappa + 4\beta^2  g^2/\Gamma}\\
&=& \gamma + \frac{4\alpha^2\lambda^2}{\kappa} +  \frac{4V_0^2}{\Gamma}  \frac{1}{ 1 + \mathcal{C}_\beta }    ,  \\
\gamma_{a_-}  &=& \gamma + \frac{4\beta^2\lambda^2}{\kappa} +  \frac{4V_0^2}{\Gamma}  \frac{\kappa}{ \kappa + 4\alpha^2 g^2/\Gamma}\\ &=&  \gamma + \frac{4\beta^2\lambda^2}{\kappa} +  \frac{4V_0^2}{\Gamma}   \frac{1}{ 1 + \mathcal{C}_\alpha }. 
\eea
\end{subequations}
This shows that the phonon linewidth has a strong dependence on the optomechanical cooperativities and thus on the optical driving strength $\alpha$ and $\beta$.

\subsection{Frequency shift}
The modification of mechanical linewidth (cooling or heating) can come with a change in the effective phonon resonant frequency, since a complex term is added to the phonon susceptibility due to the optomechanical coupling. The change in phonon frequency is related to the imaginary part of the susceptibility, and is calculated as 
\bea
\delta\omega_{a_+} (\nu) &=&    \frac{\alpha^2\lambda^2 (\nu + \Delta) }{(\nu + \Delta)^2 + \kappa^2/4} + \frac{2V_0^2}{\Gamma} \frac{2\beta^2 g^2}{\Gamma} \nonumber \\
&&  \times  \frac{\nu + \Delta}{(\nu + \Delta)^2 + (1+ \mathcal{C}_\beta)^2 \kappa^2/4 }, \\
\delta\omega_{a_-} (\nu) &=&    \frac{\beta^2\lambda^2 (\nu + \Delta) }{(\nu + \Delta)^2 + \kappa^2/4} + \frac{2V_0^2}{\Gamma} \frac{2\alpha^2 g^2}{\Gamma} \nonumber \\
&& \times  \frac{\nu + \Delta}{(\nu + \Delta)^2 +  (1+ \mathcal{C}_\alpha)^2 \kappa^2/4 }. 
\eea
The maximum frequency shift occurs when $\nu = -\Delta + \kappa/2$,  
\bea
\delta\omega_{a_+}(-\Delta +\kappa/2) &=&  \frac{\alpha^2 \lambda^2}{\kappa}  + \frac{2V_0^2}{\Gamma} \frac{\mathcal{C}_\beta}{1 + (1 + \mathcal{C}_\beta)^2 }, \\ 
\delta\omega_{a_-} (-\Delta +\kappa/2) &=&  \frac{\beta^2 \lambda^2}{\kappa}  + \frac{2V_0^2}{\Gamma} \frac{\mathcal{C}_\alpha}{1 +  (1+\mathcal{C}_\alpha)^2 }.
\eea
The second term of the frequency shift is of second order, so the main contribution comes from the first term,  which is 1/4 of the optically induced  damping rate $4\alpha^2 \lambda^2/\kappa$. In \cite{Kim2016nature}, the observed largest optomechanical damping rate is about $40\sim50~\text{kHz}$ , so the frequency shift is at most $12.5~\text{kHz}$. The cavity linewidth is $5.2~\text{MHz}$ and the detuning of the anti-Stokes line ranges from $0.2~\text{MHz}$ to $0.7~\text{MHz}$  at high power, which makes the frequency shift almost negligible.

If the phase matching condition is satisfied, then we expect $\Delta \approx - \omega_m$. In this case, when we look  at the frequency shift near original mechanical frequency $\nu \approx \omega_m $, we get $\nu + \Delta \approx 0$ and the frequency shift becomes completely negligible.

\subsection{Effective temperature}
Another important feature is the reduction in the effective temperature of the $a_-$ mode, because of coherent damping. We look at the right hand side of equation Eq.(16c) and assume that the optical noise is negligible compared to the thermal noise. We have the effective noise on $a_-$ as 
\bea
\sqrt{\gamma} a_-^{\text{in}}  - \frac{iV_0}{\sqrt{\Gamma}/2} \left( 1 - \frac{\alpha^2 g^2}{\Gamma\tilde{\kappa}/2} \right) b_+^{\text{in}} 
\eea

Assuming the initial temperature of mechanical modes are $T_{a_{\pm}}$ and $T_{b_{\pm}}$.  The effective temperature of mode $a_-$ is thus
\bea
T_{a_-}^ {\text{eff}} &=& \frac{1}{\gamma_{a_-}} \left[ \gamma T_{a_-} + \frac{V_0^2}{\Gamma/4} \abs{ 1 - \frac{\alpha^2 g^2}{\Gamma\tilde{\kappa}/2}}^2 T_{b_+} \right]  \nonumber \\
&=&  \frac{1}{\gamma_{a_-}} \left[ \gamma T_{a_-} + \frac{4V_0^2}{\Gamma} \frac{(\nu + \Delta)^2 + \kappa^2/4}{(\nu + \Delta)^2 + (\kappa/2 + 2\alpha^2 g^2/\Gamma)^2}   T_{b_+} \right] \nonumber \\
\eea
When $\nu = -\Delta$, we have
\bea
T_{a_-}^{\text{eff}}& =&  \frac{1}{\gamma_{a_-}} \left[ \gamma T_{a_-} + \frac{4V_0^2}{\Gamma} \frac{\kappa^2/4}{(\kappa/2 + 2\alpha^2 g^2/\Gamma)^2}   T_{b_+} \right] \nonumber \\
&=&\frac{1}{\gamma_{a_-}} \left[ \gamma T_{a_-} + \frac{4V_0^2}{\Gamma} \frac{1}{(1 + \mathcal{C}_\alpha)^2}   T_{b_+} \right]
\eea
We also get similar expressions for the $a_+$ mode, 
\be
T_{a_+}^{\text{eff}} =  \frac{1}{\gamma_{a_+}} \left[ \gamma T_{a_+} + \frac{4V_0^2}{\Gamma}  \frac{1}{(1 + \mathcal{C}_\beta)^2}   T_{b_-} \right]. 
\ee
When the counter-propagating pump $\beta$ is much smaller compared to the co-propagating pump $\alpha$, this effect is not so significant for the $a_+$ mode. In general, we get a correction term to the effective temperature, which roughly scales as $1/(1 + \mathcal{C}_{\alpha(\beta)} )^2$. We plot the cooperativity dependence of the linewidth and temperatures below in Fig.~3. 

\begin{figure}[!h]
\begin{center}
\includegraphics[width=.98 \columnwidth]{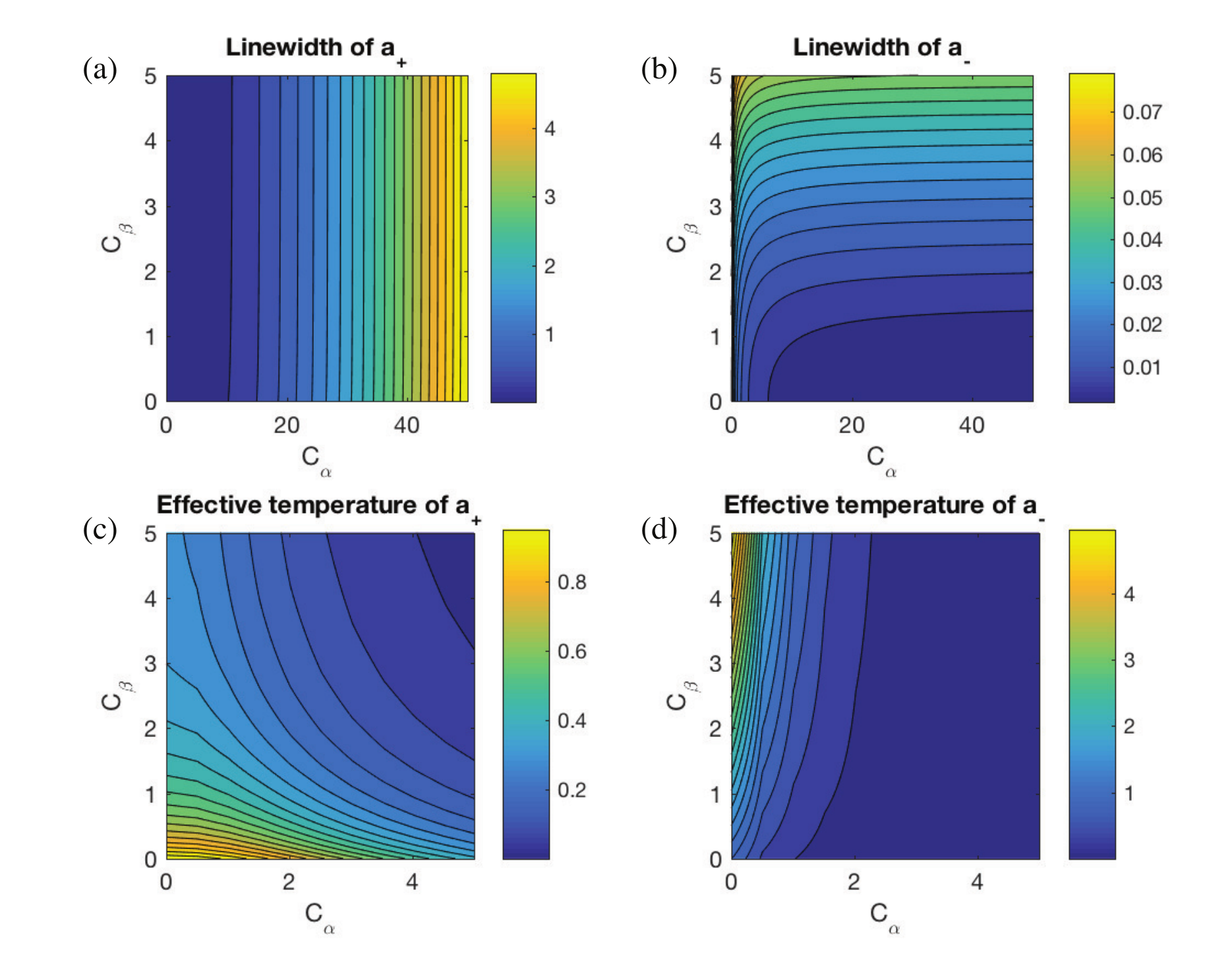}
\caption{Choosing a set of rescaled parameters of $\omega_m = 1$, $\gamma = 0.001$, $\kappa = 0.1$, $\lambda = 0.001$, $g = 0.005$, 
$\Gamma = 0.05$, $V_0 = 0.02$ and assuming that the initial thermal bath temperature $T$ is the same for all four modes,   we plot $\gamma_{a_+}/\gamma$,  $\gamma_{a_-}/\gamma$,  $T_{a_+}^{\mathrm{eff}}/T$  and  $T_{a_-}^{\mathrm{eff}}/T$ in (a)-(d) respectively.}
\label{fig:fig2}
\end{center}
\end{figure}

\subsection{Direct back-scattering corrections}
We now consider the  direct back scattering between $a_{\pm}$, with strength $V_1$, which couples the two systems in Fig.~2(b) that we have so far assumed to be independent.  Since there is a loop structure in the diagram, we look at the special case when $V_0 = 0$.  After adiabatic elimination of $c_\pm$, we have the following Heisenberg-Lagevin equations: 
\begin{subequations}
\bea
-i\nu a_+ &=&  - i\omega_{a_+} a_+ - \frac{\gamma_{a_+}}{2} a_+   - iV_1 a_-  +\sqrt{\gamma_{a_+}} a_+^{\text{in}}\nonumber  \\
-i\nu a_- &=&  - i\omega_{a_-} a_- - \frac{\gamma_{a_-}}{2} a_-   - iV_1 a_+  +\sqrt{\gamma_{a_-}} a_-^{\text{in}}  \nonumber 
\eea
\end{subequations}
where $\omega_{a_+} \approx \omega_{a_-} \approx \omega_m$,  $\gamma_{a_+} = \gamma + 4\alpha^2\lambda^2/\kappa$ and $\gamma_{a_-} = \gamma + 4\beta^2\lambda^2/\kappa$. The normal modes of these equations have resonance conditions corresponding to two poles:
\begin{equation}
\omega_{\pm} = \omega_m - i \gamma_\Sigma/2 \pm \sqrt{V_1^2 - \delta \gamma^2/4}
\end{equation}
where $\gamma_{\Sigma} = \gamma + 2(\alpha^2 + \beta^2) \lambda^2/\kappa$ is the average damping and $\delta \gamma = 4\abs{\alpha^2  - \beta^2}\lambda^2/\kappa$ is the difference in damping.  At zero power $\alpha = \beta = 0$, the two poles are split on the real axis by $\pm V_1$. As the power difference increases and $\delta\gamma > 2 V_1$, the damping rates start to differ. In an experimental that we are going to discuss in the following section,  the damping is different for all optical power, which means the prediction from the assumption of $V_1$ scattering is not applicable.

\section{Connection to experiment and parameter estimation}
\subsection{Description of the chiral phonon experiment }
A recent experimental test of this theory \cite{Kim2016nature}, considers a whispering gallery-type resonator with an intrinsic degeneracy for co-propagating and counter-propagating directions for both phonons $a_{\pm}$ and photons $c_{\pm}$.  Photons occupying the modes in the co-propagating (counter-propagating) direction can be coupled through Brillouin acousto-optic forward scattering from the co-propagating (counter-propagating) phonons.  When pumping the lower-energy optical mode, anti-Stokes scattering to the higher mode annihilates phonons in the corresponding direction and leads to unidirectional optomechanical damping \cite{Bahl2012}. In the experiment, two optical sources are tuned to the lower frequency optical mode in both the co-propagating  and counter-propagating directions, with different pump power. While one source is used as a strong pump to induce Brillouin cooling, the function of the second counter-propagating weak source is primarily to measure the modification of the high-Q phonon behavior, and the possibility of chiral behavior. 

In the experiment, a striking direction-dependence of the damping rates of the co-propagating and counter-propagating phonons was observed, as a result of the momentum conservation rules described above that underly the Brillouin scattering interaction. The experimental data points are shown in Fig.~4 below. Since the relative power of the co-propagating pump and counter-propagating probe lasers in the experiment is $\sim\,9:1$,  there is some cooling of the $a_-$ phonons as well. 

In the following, we fit the experimental results obtained in \cite{Kim2016nature}  using the theoretical model of this paper. 

\subsection{Data fitting} 
First of all, we recall the relation between the amplitudes and pump power is given by 
\bea
\alpha^2 = \fr{P_+ \kappa/2\hbar\omega}{\delta^2 + \kappa^2/4} &=& \frac{2}{\hbar\omega\kappa} \frac{P_+ (\kappa/2)^2}{\delta^2 + \kappa^2/4} =   \frac{2}{\hbar\omega\kappa} \tilde{P}_+  = \eta \tilde{P}_+   \nonumber \\
\beta^2 = \fr{P_- \kappa/2\hbar\omega}{\delta^2 + \kappa^2/4} &=&  \frac{2}{\hbar\omega\kappa}  \frac{P_- (\kappa/2)^2}{\delta^2 + \kappa^2/4} =    \frac{2}{\hbar\omega\kappa} \tilde{P}_-  = \eta \tilde{P}_- \nonumber 
\eea
where the coeffecient $\eta = 2/\hbar\omega\kappa$. 
Also, we consider the case when the counter-propagating pump is much weaker than the co-propagating pump with a ratio $P_+/P_- = \alpha^2/\beta^2 = 9.15$, then we can simplify the expression for the co-propagating phonon linewidth Eq.~(27) and get
\bea
\gamma_{a_+}  &=& \gamma + \frac{4\alpha^2\lambda^2}{\kappa} +  \frac{4V_0^2}{\Gamma}\frac{\kappa}{ \kappa + 4\beta^2 g^2/\Gamma }  \\
&\approx &  \left( \gamma + \frac{4V_0^2}{\Gamma} \right) +  \frac{4\alpha^2\lambda^2}{\kappa} \nonumber \\
&=& \gamma_{\text{eff}} + \frac{4\eta\lambda^2}{\kappa} \tilde{P}_+ 
\eea
where $\gamma_{\mathrm{eff}}$ is the sum of the intrinsic phonon linewidth and phonon scattering induced linewidth without any optical pumping. 
We fit the experimental data for co-propagating phonon linewidth with a linear model of the following form:
\be
y_+ = p_0 + p_1\cdot x, 
\ee
where $x$ represents  the detuning corrected  pump power $\tilde{P}_+ $ in units of $\mu \mathrm{W}$ and $y$ represents the linewidth $\gamma_{a_+}$ in units of $\mathrm{kHz}$. 

The asymmetry between the strength of the co-propagating and counter-propagating pump leads to a qualitatively different result for the counter-propagating phonon $a_-$, with its linewidth given by 
\bea
 \gamma_{a_-}  &=& \gamma + \frac{4\beta^2\lambda^2}{\kappa} +  \frac{4V_0^2}{\Gamma}\frac{\kappa}{ \kappa + 4\alpha^2 g^2/\Gamma }  \\
 &=& \gamma_{\mathrm{eff}}  + \frac{4\beta^2\lambda^2}{\kappa} -  \frac{4V_0^2}{\Gamma}\frac{4\alpha^2 g^2/\Gamma}{ \kappa + 4\alpha^2 g^2/\Gamma } 
\eea
The $\alpha^2$ term can be large, and it is this term that leads to a substantial linewidth reduction for the counter-propagating phonon as we increase the co-propagating pump power. 

For the low power data, the ratio $r = \alpha^2/\beta^2$ is fixed at $9.15$, but for the high power data, $\beta$ itself is fixed at certain value $\beta_0$. For data taken under different conditions, we may have a discontinuous change in the counter-propagating pump power. So we can use a piecewise function to describe the relation between the two pumps as:
\be
\beta^2 = (\alpha\leq \alpha_0) ~\alpha^2/r  + (\alpha>\alpha_0)~ s \alpha_0^2. 
\ee
We can use a corresponding piecewise function to describe the data for the broadened linewidth at low power:
\be
y_- = q_0 + q_1 \cdot x, \quad \text{for} ~(x<P_0) . 
\ee

\begin{figure}[!ht]
\begin{center}
\includegraphics[width=.7\columnwidth]{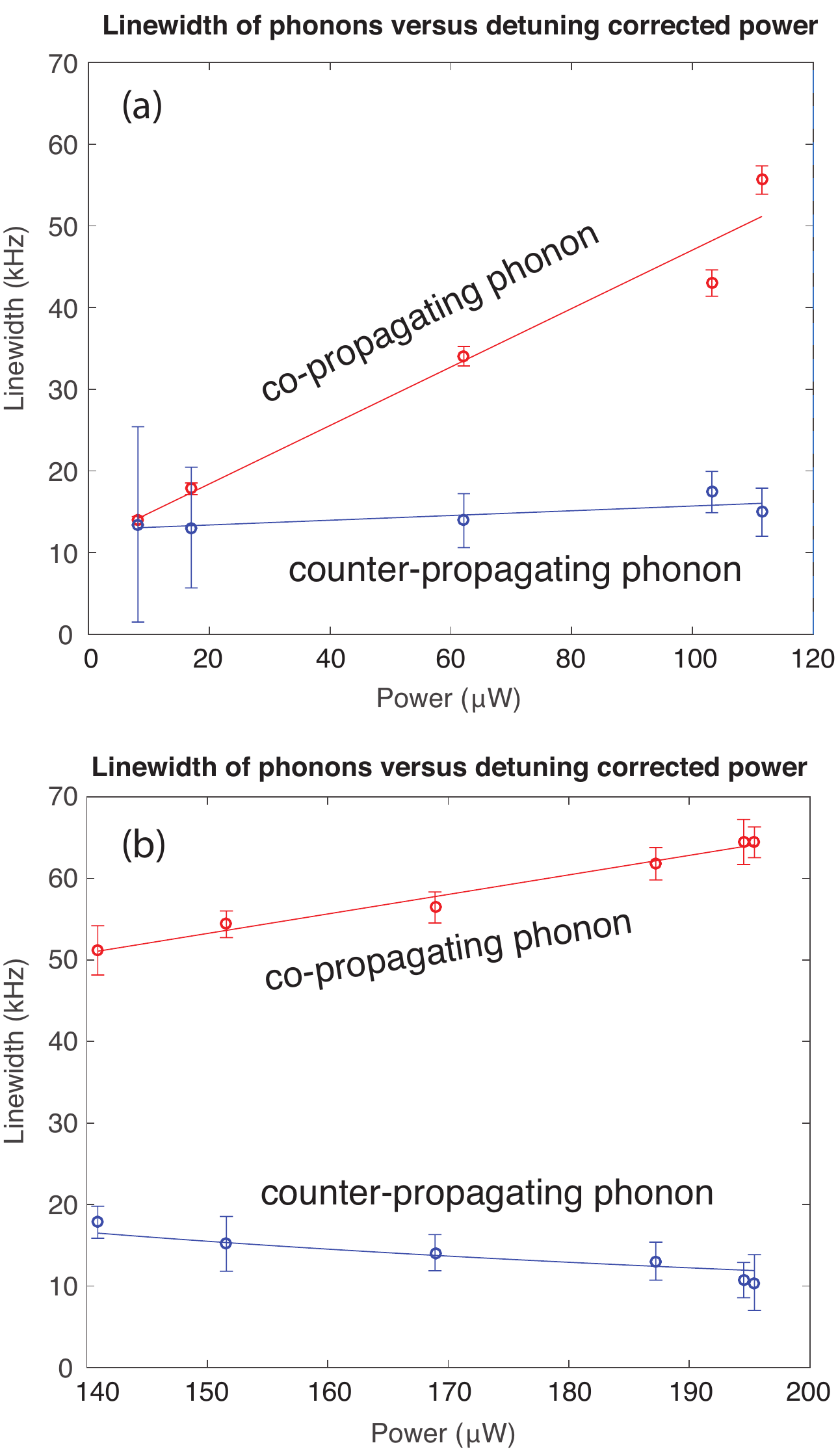}
\caption{Red is for co-propagating phonon and blue is for counter-propagating phonon. (a) is for low power data and (b) is for high power data. Experimental data points (in circles) are taken from \cite{Kim2016nature} and solid lines are fits to the data points.}
\label{fig:fig3}
\end{center}
\end{figure}

Least square fitting results for low power data and high power data are shown in Fig. 4, and fit values are shown in Table I. For the forward propagating direction, the parameter $\vec{p} = (p_0, p_1)$ is quite consistent between low power and high power data. For the counter-propagating direction, the situation is more complicated.  From the lower power data, we can only get enough information about $q_1$ and $q_2$. At high power, the fitting would nominally take a form (based upon Eq. (44))
\be
y_+ = q_2 + \frac{q_3}{q_4 + x} 
\ee
However, fitting $q_3$, $q_4$ as well as $\beta_0$, leads to substantial correlation and likely overfitting. Instead, we focus on a simpler model at high power to capture the reduction of linewidth. Specifically, we fix $q_2 = q_0 + \alpha_0 q_1$ from the low power data, and set $q_4 = 0$. 

\begin{table}[!ht]
\caption{Fitting results (with $2\sigma$ error) for low power and high power data  }
\begin{center}
\begin{tabular}{l|c|c|c|c}
\hline
parameter (units) \hspace{4em}  & \hspace{.5em}  value for low power data  \hspace{.5em}   &  \hspace{.5em} value for high power data  \hspace{.5em}   \\ \hline
$p_0 = \gamma_{\mathrm{eff}}~(\mathrm{kHz})$  & 11.25(9.83) & 17.31(7.98) \\ \hline
$p_1 = 4\eta\lambda^2/\kappa~(\mathrm{kHz}/\mu\mathrm{W}) $ & 0.36(0.13) & 0.24(0.05) \\ \hline 
$q_0 = \gamma ~(\mathrm{kHz})$ & 12.80(3.02) & NA \\ \hline
$q_1  = 4\eta\lambda^2/\kappa r~(\mathrm{kHz}/\mu\mathrm{W}) $ & 0.029(0.041)  & NA \\ \hline 
$q_3 = \kappa V_0^2/\eta g^2~(\mathrm{kHz} \cdot \mu\mathrm{W})$ & NA  & 2324(194)   \\ \hline
\end{tabular}
\end{center}
\label{table1}
\end{table}

\section{Conclusion}
In this article, we present a model for optomechanically induced chiral phonon behavior. We show that in Brillouin sideband cooling experiments on traveling phonon populations, the linewidth of co-propagating phonons is increased by the optomechanical interaction with optical driving fields, while the linewidth of counter-propagating phonons is decreased at the same time. We also predict the effective temperature of counter-propagating phonon will decrease by increasing the driving fields, which is in contrast to conventional optomechanical cooling where phonon linewidth and effective temperature move in different directions. This model is able to explain the chiral behavior of phonon transport observed in a recent experiment \cite{Kim2016nature}.

\section{Acknowledgements}
We thank Thomas Purdy for helpful discussions.  Funding for this research was provided through the National Science Foundation (NSF), Air Force Office for Scientific Research (AFOSR), the Office of Naval Research (ONR), and DARPA MTO.

\bibliography{chirality_v13}

\providecommand{\noopsort}[1]{}\providecommand{\singleletter}[1]{#1}%
\begin{thebibliography}{32}%
\makeatletter
\providecommand \@ifxundefined [1]{%
 \@ifx{#1\undefined}
}%
\providecommand \@ifnum [1]{%
 \ifnum #1\expandafter \@firstoftwo
 \else \expandafter \@secondoftwo
 \fi
}%
\providecommand \@ifx [1]{%
 \ifx #1\expandafter \@firstoftwo
 \else \expandafter \@secondoftwo
 \fi
}%
\providecommand \natexlab [1]{#1}%
\providecommand \enquote  [1]{``#1''}%
\providecommand \bibnamefont  [1]{#1}%
\providecommand \bibfnamefont [1]{#1}%
\providecommand \citenamefont [1]{#1}%
\providecommand \href@noop [0]{\@secondoftwo}%
\providecommand \href [0]{\begingroup \@sanitize@url \@href}%
\providecommand \@href[1]{\@@startlink{#1}\@@href}%
\providecommand \@@href[1]{\endgroup#1\@@endlink}%
\providecommand \@sanitize@url [0]{\catcode `\\12\catcode `\$12\catcode
  `\&12\catcode `\#12\catcode `\^12\catcode `\_12\catcode `\%12\relax}%
\providecommand \@@startlink[1]{}%
\providecommand \@@endlink[0]{}%
\providecommand \url  [0]{\begingroup\@sanitize@url \@url }%
\providecommand \@url [1]{\endgroup\@href {#1}{\urlprefix }}%
\providecommand \urlprefix  [0]{URL }%
\providecommand \Eprint [0]{\href }%
\providecommand \doibase [0]{http://dx.doi.org/}%
\providecommand \selectlanguage [0]{\@gobble}%
\providecommand \bibinfo  [0]{\@secondoftwo}%
\providecommand \bibfield  [0]{\@secondoftwo}%
\providecommand \translation [1]{[#1]}%
\providecommand \BibitemOpen [0]{}%
\providecommand \bibitemStop [0]{}%
\providecommand \bibitemNoStop [0]{.\EOS\space}%
\providecommand \EOS [0]{\spacefactor3000\relax}%
\providecommand \BibitemShut  [1]{\csname bibitem#1\endcsname}%
\let\auto@bib@innerbib\@empty
\bibitem [{\citenamefont {Caves}(1980)}]{Caves1980}%
  \BibitemOpen
  \bibfield  {author} {\bibinfo {author} {\bibfnamefont {C.~M.}\ \bibnamefont
  {Caves}},\ }\href {\doibase 10.1103/PhysRevLett.45.75} {\bibfield  {journal}
  {\bibinfo  {journal} {Phys. Rev. Lett.}\ }\textbf {\bibinfo {volume} {45}},\
  \bibinfo {pages} {75} (\bibinfo {year} {1980})}\BibitemShut {NoStop}%
\bibitem [{\citenamefont {Braginsky}\ \emph {et~al.}(2001)\citenamefont
  {Braginsky}, \citenamefont {Strigin},\ and\ \citenamefont
  {Vyatchanin}}]{Braginsky2001}%
  \BibitemOpen
  \bibfield  {author} {\bibinfo {author} {\bibfnamefont {V.}~\bibnamefont
  {Braginsky}}, \bibinfo {author} {\bibfnamefont {S.}~\bibnamefont {Strigin}},
  \ and\ \bibinfo {author} {\bibfnamefont {S.}~\bibnamefont {Vyatchanin}},\
  }\href {\doibase http://dx.doi.org/10.1016/S0375-9601(01)00510-2} {\bibfield
  {journal} {\bibinfo  {journal} {Physics Letters A}\ }\textbf {\bibinfo
  {volume} {287}},\ \bibinfo {pages} {331 } (\bibinfo {year}
  {2001})}\BibitemShut {NoStop}%
\bibitem [{\citenamefont {Braginsky}\ \emph {et~al.}(2002)\citenamefont
  {Braginsky}, \citenamefont {Strigin},\ and\ \citenamefont
  {Vyatchanin}}]{Braginsky2002}%
  \BibitemOpen
  \bibfield  {author} {\bibinfo {author} {\bibfnamefont {V.}~\bibnamefont
  {Braginsky}}, \bibinfo {author} {\bibfnamefont {S.}~\bibnamefont {Strigin}},
  \ and\ \bibinfo {author} {\bibfnamefont {S.}~\bibnamefont {Vyatchanin}},\
  }\href {\doibase http://dx.doi.org/10.1016/S0375-9601(02)01357-9} {\bibfield
  {journal} {\bibinfo  {journal} {Physics Letters A}\ }\textbf {\bibinfo
  {volume} {305}},\ \bibinfo {pages} {111 } (\bibinfo {year}
  {2002})}\BibitemShut {NoStop}%
\bibitem [{\citenamefont {Kippenberg}\ \emph {et~al.}(2005)\citenamefont
  {Kippenberg}, \citenamefont {Rokhsari}, \citenamefont {Carmon}, \citenamefont
  {Scherer},\ and\ \citenamefont {Vahala}}]{Kippenberg2005}%
  \BibitemOpen
  \bibfield  {author} {\bibinfo {author} {\bibfnamefont {T.~J.}\ \bibnamefont
  {Kippenberg}}, \bibinfo {author} {\bibfnamefont {H.}~\bibnamefont
  {Rokhsari}}, \bibinfo {author} {\bibfnamefont {T.}~\bibnamefont {Carmon}},
  \bibinfo {author} {\bibfnamefont {A.}~\bibnamefont {Scherer}}, \ and\
  \bibinfo {author} {\bibfnamefont {K.~J.}\ \bibnamefont {Vahala}},\ }\href
  {\doibase 10.1103/PhysRevLett.95.033901} {\bibfield  {journal} {\bibinfo
  {journal} {Phys. Rev. Lett.}\ }\textbf {\bibinfo {volume} {95}},\ \bibinfo
  {pages} {033901} (\bibinfo {year} {2005})}\BibitemShut {NoStop}%
\bibitem [{\citenamefont {Kippenberg}\ and\ \citenamefont
  {Vahala}(2008)}]{Kippenberg2008}%
  \BibitemOpen
  \bibfield  {author} {\bibinfo {author} {\bibfnamefont {T.~J.}\ \bibnamefont
  {Kippenberg}}\ and\ \bibinfo {author} {\bibfnamefont {K.~J.}\ \bibnamefont
  {Vahala}},\ }\href {http://www.sciencemag.org/content/321/5893/1172.full}
  {\bibfield  {journal} {\bibinfo  {journal} {Science}\ }\textbf {\bibinfo
  {volume} {321}},\ \bibinfo {pages} {1172} (\bibinfo {year}
  {2008})}\BibitemShut {NoStop}%
\bibitem [{\citenamefont {Marquardt}\ and\ \citenamefont
  {Girvin}(2009)}]{Marquardt2009}%
  \BibitemOpen
  \bibfield  {author} {\bibinfo {author} {\bibfnamefont {F.}~\bibnamefont
  {Marquardt}}\ and\ \bibinfo {author} {\bibfnamefont {S.~M.}\ \bibnamefont
  {Girvin}},\ }\href {http://physics.aps.org/articles/v2/40} {\bibfield
  {journal} {\bibinfo  {journal} {Physics}\ }\textbf {\bibinfo {volume} {2}},\
  \bibinfo {pages} {40} (\bibinfo {year} {2009})}\BibitemShut {NoStop}%
\bibitem [{\citenamefont {Aspelmeyer}\ \emph {et~al.}(2014)\citenamefont
  {Aspelmeyer}, \citenamefont {Kippenberg},\ and\ \citenamefont
  {Marquardt}}]{Aspelmeyer2014}%
  \BibitemOpen
  \bibfield  {author} {\bibinfo {author} {\bibfnamefont {M.}~\bibnamefont
  {Aspelmeyer}}, \bibinfo {author} {\bibfnamefont {T.~J.}\ \bibnamefont
  {Kippenberg}}, \ and\ \bibinfo {author} {\bibfnamefont {F.}~\bibnamefont
  {Marquardt}},\ }\href {\doibase 10.1103/RevModPhys.86.1391} {\bibfield
  {journal} {\bibinfo  {journal} {Rev. Mod. Phys.}\ }\textbf {\bibinfo {volume}
  {86}},\ \bibinfo {pages} {1391} (\bibinfo {year} {2014})}\BibitemShut
  {NoStop}%
\bibitem [{\citenamefont {Wilson-Rae}\ \emph {et~al.}(2007)\citenamefont
  {Wilson-Rae}, \citenamefont {Nooshi}, \citenamefont {Zwerger},\ and\
  \citenamefont {Kippenberg}}]{WilsonRae2007}%
  \BibitemOpen
  \bibfield  {author} {\bibinfo {author} {\bibfnamefont {I.}~\bibnamefont
  {Wilson-Rae}}, \bibinfo {author} {\bibfnamefont {N.}~\bibnamefont {Nooshi}},
  \bibinfo {author} {\bibfnamefont {W.}~\bibnamefont {Zwerger}}, \ and\
  \bibinfo {author} {\bibfnamefont {T.~J.}\ \bibnamefont {Kippenberg}},\ }\href
  {\doibase 10.1103/PhysRevLett.99.093901} {\bibfield  {journal} {\bibinfo
  {journal} {Phys. Rev. Lett.}\ }\textbf {\bibinfo {volume} {99}},\ \bibinfo
  {pages} {093901} (\bibinfo {year} {2007})}\BibitemShut {NoStop}%
\bibitem [{\citenamefont {Teufel}\ \emph {et~al.}(2011)\citenamefont {Teufel},
  \citenamefont {Donner}, \citenamefont {Li}, \citenamefont {Harlow},
  \citenamefont {Allman}, \citenamefont {Cicak}, \citenamefont {Sirois},
  \citenamefont {Whittaker}, \citenamefont {Lehnert},\ and\ \citenamefont
  {Simmonds}}]{Teufel2011}%
  \BibitemOpen
  \bibfield  {author} {\bibinfo {author} {\bibfnamefont {J.~D.}\ \bibnamefont
  {Teufel}}, \bibinfo {author} {\bibfnamefont {T.}~\bibnamefont {Donner}},
  \bibinfo {author} {\bibfnamefont {D.}~\bibnamefont {Li}}, \bibinfo {author}
  {\bibfnamefont {J.~W.}\ \bibnamefont {Harlow}}, \bibinfo {author}
  {\bibfnamefont {M.~S.}\ \bibnamefont {Allman}}, \bibinfo {author}
  {\bibfnamefont {K.}~\bibnamefont {Cicak}}, \bibinfo {author} {\bibfnamefont
  {A.~J.}\ \bibnamefont {Sirois}}, \bibinfo {author} {\bibfnamefont {J.~D.}\
  \bibnamefont {Whittaker}}, \bibinfo {author} {\bibfnamefont {K.~W.}\
  \bibnamefont {Lehnert}}, \ and\ \bibinfo {author} {\bibfnamefont {R.~W.}\
  \bibnamefont {Simmonds}},\ }\href {http://dx.doi.org/10.1038/nature10261}
  {\bibfield  {journal} {\bibinfo  {journal} {Nature}\ }\textbf {\bibinfo
  {volume} {475}},\ \bibinfo {pages} {359} (\bibinfo {year}
  {2011})}\BibitemShut {NoStop}%
\bibitem [{\citenamefont {Chan}\ \emph {et~al.}(2011)\citenamefont {Chan},
  \citenamefont {Alegre}, \citenamefont {Safavi-Naeini}, \citenamefont {Hill},
  \citenamefont {Krause}, \citenamefont {Groblacher}, \citenamefont
  {Aspelmeyer},\ and\ \citenamefont {Painter}}]{Chan2011}%
  \BibitemOpen
  \bibfield  {author} {\bibinfo {author} {\bibfnamefont {J.}~\bibnamefont
  {Chan}}, \bibinfo {author} {\bibfnamefont {T.~P.~M.}\ \bibnamefont {Alegre}},
  \bibinfo {author} {\bibfnamefont {A.~H.}\ \bibnamefont {Safavi-Naeini}},
  \bibinfo {author} {\bibfnamefont {J.~T.}\ \bibnamefont {Hill}}, \bibinfo
  {author} {\bibfnamefont {A.}~\bibnamefont {Krause}}, \bibinfo {author}
  {\bibfnamefont {S.}~\bibnamefont {Groblacher}}, \bibinfo {author}
  {\bibfnamefont {M.}~\bibnamefont {Aspelmeyer}}, \ and\ \bibinfo {author}
  {\bibfnamefont {O.}~\bibnamefont {Painter}},\ }\href
  {http://dx.doi.org/10.1038/nature10461} {\bibfield  {journal} {\bibinfo
  {journal} {Nature}\ }\textbf {\bibinfo {volume} {478}},\ \bibinfo {pages}
  {89} (\bibinfo {year} {2011})}\BibitemShut {NoStop}%
\bibitem [{\citenamefont {Brooks}\ \emph {et~al.}(2012)\citenamefont {Brooks},
  \citenamefont {Botter}, \citenamefont {Schreppler}, \citenamefont {Purdy},
  \citenamefont {Brahms},\ and\ \citenamefont {Stamper-Kurn}}]{Brooks2012}%
  \BibitemOpen
  \bibfield  {author} {\bibinfo {author} {\bibfnamefont {D.~W.~C.}\
  \bibnamefont {Brooks}}, \bibinfo {author} {\bibfnamefont {T.}~\bibnamefont
  {Botter}}, \bibinfo {author} {\bibfnamefont {S.}~\bibnamefont {Schreppler}},
  \bibinfo {author} {\bibfnamefont {T.~P.}\ \bibnamefont {Purdy}}, \bibinfo
  {author} {\bibfnamefont {N.}~\bibnamefont {Brahms}}, \ and\ \bibinfo {author}
  {\bibfnamefont {D.~M.}\ \bibnamefont {Stamper-Kurn}},\ }\href
  {http://dx.doi.org/10.1038/nature11325} {\bibfield  {journal} {\bibinfo
  {journal} {Nature}\ }\textbf {\bibinfo {volume} {488}},\ \bibinfo {pages}
  {476} (\bibinfo {year} {2012})}\BibitemShut {NoStop}%
\bibitem [{\citenamefont {Safavi-Naeini}\ \emph {et~al.}(2013)\citenamefont
  {Safavi-Naeini}, \citenamefont {Groblacher}, \citenamefont {Hill},
  \citenamefont {Chan}, \citenamefont {Aspelmeyer},\ and\ \citenamefont
  {Painter}}]{SafaviNaeini2013}%
  \BibitemOpen
  \bibfield  {author} {\bibinfo {author} {\bibfnamefont {A.~H.}\ \bibnamefont
  {Safavi-Naeini}}, \bibinfo {author} {\bibfnamefont {S.}~\bibnamefont
  {Groblacher}}, \bibinfo {author} {\bibfnamefont {J.~T.}\ \bibnamefont
  {Hill}}, \bibinfo {author} {\bibfnamefont {J.}~\bibnamefont {Chan}}, \bibinfo
  {author} {\bibfnamefont {M.}~\bibnamefont {Aspelmeyer}}, \ and\ \bibinfo
  {author} {\bibfnamefont {O.}~\bibnamefont {Painter}},\ }\href
  {http://dx.doi.org/10.1038/nature12307} {\bibfield  {journal} {\bibinfo
  {journal} {Nature}\ }\textbf {\bibinfo {volume} {500}},\ \bibinfo {pages}
  {185} (\bibinfo {year} {2013})}\BibitemShut {NoStop}%
\bibitem [{\citenamefont {Purdy}\ \emph {et~al.}(2013)\citenamefont {Purdy},
  \citenamefont {Yu}, \citenamefont {Peterson}, \citenamefont {Kampel},\ and\
  \citenamefont {Regal}}]{Purdy2013}%
  \BibitemOpen
  \bibfield  {author} {\bibinfo {author} {\bibfnamefont {T.~P.}\ \bibnamefont
  {Purdy}}, \bibinfo {author} {\bibfnamefont {P.-L.}\ \bibnamefont {Yu}},
  \bibinfo {author} {\bibfnamefont {R.~W.}\ \bibnamefont {Peterson}}, \bibinfo
  {author} {\bibfnamefont {N.~S.}\ \bibnamefont {Kampel}}, \ and\ \bibinfo
  {author} {\bibfnamefont {C.~A.}\ \bibnamefont {Regal}},\ }\href {\doibase
  10.1103/PhysRevX.3.031012} {\bibfield  {journal} {\bibinfo  {journal} {Phys.
  Rev. X}\ }\textbf {\bibinfo {volume} {3}},\ \bibinfo {pages} {031012}
  (\bibinfo {year} {2013})}\BibitemShut {NoStop}%
\bibitem [{\citenamefont {Nunnenkamp}\ \emph {et~al.}(2011)\citenamefont
  {Nunnenkamp}, \citenamefont {B\o{}rkje},\ and\ \citenamefont
  {Girvin}}]{Nunnenkamp11}%
  \BibitemOpen
  \bibfield  {author} {\bibinfo {author} {\bibfnamefont {A.}~\bibnamefont
  {Nunnenkamp}}, \bibinfo {author} {\bibfnamefont {K.}~\bibnamefont
  {B\o{}rkje}}, \ and\ \bibinfo {author} {\bibfnamefont {S.~M.}\ \bibnamefont
  {Girvin}},\ }\href {\doibase 10.1103/PhysRevLett.107.063602} {\bibfield
  {journal} {\bibinfo  {journal} {Phys. Rev. Lett.}\ }\textbf {\bibinfo
  {volume} {107}},\ \bibinfo {pages} {063602} (\bibinfo {year}
  {2011})}\BibitemShut {NoStop}%
\bibitem [{\citenamefont {Rabl}(2011)}]{Rabl2011}%
  \BibitemOpen
  \bibfield  {author} {\bibinfo {author} {\bibfnamefont {P.}~\bibnamefont
  {Rabl}},\ }\href {\doibase 10.1103/PhysRevLett.107.063601} {\bibfield
  {journal} {\bibinfo  {journal} {Phys. Rev. Lett.}\ }\textbf {\bibinfo
  {volume} {107}},\ \bibinfo {pages} {063601} (\bibinfo {year}
  {2011})}\BibitemShut {NoStop}%
\bibitem [{\citenamefont {Kronwald}\ and\ \citenamefont
  {Marquardt}(2013)}]{Kronwald2013}%
  \BibitemOpen
  \bibfield  {author} {\bibinfo {author} {\bibfnamefont {A.}~\bibnamefont
  {Kronwald}}\ and\ \bibinfo {author} {\bibfnamefont {F.}~\bibnamefont
  {Marquardt}},\ }\href {\doibase 10.1103/PhysRevLett.111.133601} {\bibfield
  {journal} {\bibinfo  {journal} {Phys. Rev. Lett.}\ }\textbf {\bibinfo
  {volume} {111}},\ \bibinfo {pages} {133601} (\bibinfo {year}
  {2013})}\BibitemShut {NoStop}%
\bibitem [{\citenamefont {Lemonde}\ \emph {et~al.}(2013)\citenamefont
  {Lemonde}, \citenamefont {Didier},\ and\ \citenamefont
  {Clerk}}]{Lemonde2013}%
  \BibitemOpen
  \bibfield  {author} {\bibinfo {author} {\bibfnamefont {M.-A.}\ \bibnamefont
  {Lemonde}}, \bibinfo {author} {\bibfnamefont {N.}~\bibnamefont {Didier}}, \
  and\ \bibinfo {author} {\bibfnamefont {A.~A.}\ \bibnamefont {Clerk}},\ }\href
  {\doibase 10.1103/PhysRevLett.111.053602} {\bibfield  {journal} {\bibinfo
  {journal} {Phys. Rev. Lett.}\ }\textbf {\bibinfo {volume} {111}},\ \bibinfo
  {pages} {053602} (\bibinfo {year} {2013})}\BibitemShut {NoStop}%
\bibitem [{\citenamefont {B\o{}rkje}\ \emph {et~al.}(2013)\citenamefont
  {B\o{}rkje}, \citenamefont {Nunnenkamp}, \citenamefont {Teufel},\ and\
  \citenamefont {Girvin}}]{Borkje2013}%
  \BibitemOpen
  \bibfield  {author} {\bibinfo {author} {\bibfnamefont {K.}~\bibnamefont
  {B\o{}rkje}}, \bibinfo {author} {\bibfnamefont {A.}~\bibnamefont
  {Nunnenkamp}}, \bibinfo {author} {\bibfnamefont {J.~D.}\ \bibnamefont
  {Teufel}}, \ and\ \bibinfo {author} {\bibfnamefont {S.~M.}\ \bibnamefont
  {Girvin}},\ }\href {\doibase 10.1103/PhysRevLett.111.053603} {\bibfield
  {journal} {\bibinfo  {journal} {Phys. Rev. Lett.}\ }\textbf {\bibinfo
  {volume} {111}},\ \bibinfo {pages} {053603} (\bibinfo {year}
  {2013})}\BibitemShut {NoStop}%
\bibitem [{\citenamefont {Xu}\ \emph {et~al.}(2015)\citenamefont {Xu},
  \citenamefont {Gullans},\ and\ \citenamefont {Taylor}}]{Xu2015}%
  \BibitemOpen
  \bibfield  {author} {\bibinfo {author} {\bibfnamefont {X.}~\bibnamefont
  {Xu}}, \bibinfo {author} {\bibfnamefont {M.}~\bibnamefont {Gullans}}, \ and\
  \bibinfo {author} {\bibfnamefont {J.~M.}\ \bibnamefont {Taylor}},\ }\href
  {\doibase 10.1103/PhysRevA.91.013818} {\bibfield  {journal} {\bibinfo
  {journal} {Phys. Rev. A}\ }\textbf {\bibinfo {volume} {91}},\ \bibinfo
  {pages} {013818} (\bibinfo {year} {2015})}\BibitemShut {NoStop}%
\bibitem [{\citenamefont {Grudinin}\ \emph {et~al.}(2009)\citenamefont
  {Grudinin}, \citenamefont {Matsko},\ and\ \citenamefont
  {Maleki}}]{Grudinin2009}%
  \BibitemOpen
  \bibfield  {author} {\bibinfo {author} {\bibfnamefont {I.~S.}\ \bibnamefont
  {Grudinin}}, \bibinfo {author} {\bibfnamefont {A.~B.}\ \bibnamefont
  {Matsko}}, \ and\ \bibinfo {author} {\bibfnamefont {L.}~\bibnamefont
  {Maleki}},\ }\href {\doibase 10.1103/PhysRevLett.102.043902} {\bibfield
  {journal} {\bibinfo  {journal} {Phys. Rev. Lett.}\ }\textbf {\bibinfo
  {volume} {102}},\ \bibinfo {pages} {043902} (\bibinfo {year}
  {2009})}\BibitemShut {NoStop}%
\bibitem [{\citenamefont {Grudinin}\ \emph {et~al.}(2010)\citenamefont
  {Grudinin}, \citenamefont {Lee}, \citenamefont {Painter},\ and\ \citenamefont
  {Vahala}}]{Grudinin2010}%
  \BibitemOpen
  \bibfield  {author} {\bibinfo {author} {\bibfnamefont {I.~S.}\ \bibnamefont
  {Grudinin}}, \bibinfo {author} {\bibfnamefont {H.}~\bibnamefont {Lee}},
  \bibinfo {author} {\bibfnamefont {O.}~\bibnamefont {Painter}}, \ and\
  \bibinfo {author} {\bibfnamefont {K.~J.}\ \bibnamefont {Vahala}},\ }\href
  {\doibase 10.1103/PhysRevLett.104.083901} {\bibfield  {journal} {\bibinfo
  {journal} {Phys. Rev. Lett.}\ }\textbf {\bibinfo {volume} {104}},\ \bibinfo
  {pages} {083901} (\bibinfo {year} {2010})}\BibitemShut {NoStop}%
\bibitem [{\citenamefont {Bahl}\ \emph {et~al.}(2013)\citenamefont {Bahl},
  \citenamefont {Kim}, \citenamefont {Lee}, \citenamefont {Liu}, \citenamefont
  {Fan},\ and\ \citenamefont {Carmon}}]{Bahl2013}%
  \BibitemOpen
  \bibfield  {author} {\bibinfo {author} {\bibfnamefont {G.}~\bibnamefont
  {Bahl}}, \bibinfo {author} {\bibfnamefont {K.~H.}\ \bibnamefont {Kim}},
  \bibinfo {author} {\bibfnamefont {W.}~\bibnamefont {Lee}}, \bibinfo {author}
  {\bibfnamefont {J.}~\bibnamefont {Liu}}, \bibinfo {author} {\bibfnamefont
  {X.}~\bibnamefont {Fan}}, \ and\ \bibinfo {author} {\bibfnamefont
  {T.}~\bibnamefont {Carmon}},\ }\href {http://dx.doi.org/10.1038/ncomms2994}
  {\bibfield  {journal} {\bibinfo  {journal} {Nat Commun}\ }\textbf {\bibinfo
  {volume} {4}} (\bibinfo {year} {2013})}\BibitemShut {NoStop}%
\bibitem [{\citenamefont {Bahl}\ \emph {et~al.}(2012)\citenamefont {Bahl},
  \citenamefont {Tomes}, \citenamefont {Marquardt},\ and\ \citenamefont
  {Carmon}}]{Bahl2012}%
  \BibitemOpen
  \bibfield  {author} {\bibinfo {author} {\bibfnamefont {G.}~\bibnamefont
  {Bahl}}, \bibinfo {author} {\bibfnamefont {M.}~\bibnamefont {Tomes}},
  \bibinfo {author} {\bibfnamefont {F.}~\bibnamefont {Marquardt}}, \ and\
  \bibinfo {author} {\bibfnamefont {T.}~\bibnamefont {Carmon}},\ }\href
  {http://dx.doi.org/10.1038/nphys2206} {\bibfield  {journal} {\bibinfo
  {journal} {Nat Phys}\ }\textbf {\bibinfo {volume} {8}},\ \bibinfo {pages}
  {203} (\bibinfo {year} {2012})}\BibitemShut {NoStop}%
\bibitem [{\citenamefont {Kim}\ \emph {et~al.}(2016)\citenamefont {Kim},
  \citenamefont {Xu}, \citenamefont {Taylor},\ and\ \citenamefont
  {Bahl}}]{Kim2016nature}%
  \BibitemOpen
  \bibfield  {author} {\bibinfo {author} {\bibfnamefont {S.}~\bibnamefont
  {Kim}}, \bibinfo {author} {\bibfnamefont {X.}~\bibnamefont {Xu}}, \bibinfo
  {author} {\bibfnamefont {J.~M.}\ \bibnamefont {Taylor}}, \ and\ \bibinfo
  {author} {\bibfnamefont {G.}~\bibnamefont {Bahl}},\ }\href@noop {} {\enquote
  {\bibinfo {title} {Dynamically induced robust phonon transport and chiral
  cooling in an optomechanical system},}\ }\bibinfo {howpublished} {e-print
  arXiv:1609.08674} (\bibinfo {year} {2016})\BibitemShut {NoStop}%
\bibitem [{\citenamefont {Pichler}\ \emph {et~al.}(2015)\citenamefont
  {Pichler}, \citenamefont {Ramos}, \citenamefont {Daley},\ and\ \citenamefont
  {Zoller}}]{Pichler15}%
  \BibitemOpen
  \bibfield  {author} {\bibinfo {author} {\bibfnamefont {H.}~\bibnamefont
  {Pichler}}, \bibinfo {author} {\bibfnamefont {T.}~\bibnamefont {Ramos}},
  \bibinfo {author} {\bibfnamefont {A.~J.}\ \bibnamefont {Daley}}, \ and\
  \bibinfo {author} {\bibfnamefont {P.}~\bibnamefont {Zoller}},\ }\href
  {\doibase 10.1103/PhysRevA.91.042116} {\bibfield  {journal} {\bibinfo
  {journal} {Phys. Rev. A}\ }\textbf {\bibinfo {volume} {91}},\ \bibinfo
  {pages} {042116} (\bibinfo {year} {2015})}\BibitemShut {NoStop}%
\bibitem [{\citenamefont {Ramos}\ \emph {et~al.}(2016)\citenamefont {Ramos},
  \citenamefont {Vermersch}, \citenamefont {Hauke}, \citenamefont {Pichler},\
  and\ \citenamefont {Zoller}}]{Ramos2016}%
  \BibitemOpen
  \bibfield  {author} {\bibinfo {author} {\bibfnamefont {T.}~\bibnamefont
  {Ramos}}, \bibinfo {author} {\bibfnamefont {B.}~\bibnamefont {Vermersch}},
  \bibinfo {author} {\bibfnamefont {P.}~\bibnamefont {Hauke}}, \bibinfo
  {author} {\bibfnamefont {H.}~\bibnamefont {Pichler}}, \ and\ \bibinfo
  {author} {\bibfnamefont {P.}~\bibnamefont {Zoller}},\ }\href {\doibase
  10.1103/PhysRevA.93.062104} {\bibfield  {journal} {\bibinfo  {journal} {Phys.
  Rev. A}\ }\textbf {\bibinfo {volume} {93}},\ \bibinfo {pages} {062104}
  (\bibinfo {year} {2016})}\BibitemShut {NoStop}%
\bibitem [{\citenamefont {Vermersch}\ \emph {et~al.}(2016)\citenamefont
  {Vermersch}, \citenamefont {Ramos}, \citenamefont {Hauke},\ and\
  \citenamefont {Zoller}}]{Vermersch2016}%
  \BibitemOpen
  \bibfield  {author} {\bibinfo {author} {\bibfnamefont {B.}~\bibnamefont
  {Vermersch}}, \bibinfo {author} {\bibfnamefont {T.}~\bibnamefont {Ramos}},
  \bibinfo {author} {\bibfnamefont {P.}~\bibnamefont {Hauke}}, \ and\ \bibinfo
  {author} {\bibfnamefont {P.}~\bibnamefont {Zoller}},\ }\href {\doibase
  10.1103/PhysRevA.93.063830} {\bibfield  {journal} {\bibinfo  {journal} {Phys.
  Rev. A}\ }\textbf {\bibinfo {volume} {93}},\ \bibinfo {pages} {063830}
  (\bibinfo {year} {2016})}\BibitemShut {NoStop}%
\bibitem [{\citenamefont {Chan}\ \emph {et~al.}(2012)\citenamefont {Chan},
  \citenamefont {Safavi-Naeini}, \citenamefont {Hill}, \citenamefont
  {Meenehan},\ and\ \citenamefont {Painter}}]{Chan2012}%
  \BibitemOpen
  \bibfield  {author} {\bibinfo {author} {\bibfnamefont {J.}~\bibnamefont
  {Chan}}, \bibinfo {author} {\bibfnamefont {A.~H.}\ \bibnamefont
  {Safavi-Naeini}}, \bibinfo {author} {\bibfnamefont {J.~T.}\ \bibnamefont
  {Hill}}, \bibinfo {author} {\bibfnamefont {S.}~\bibnamefont {Meenehan}}, \
  and\ \bibinfo {author} {\bibfnamefont {O.}~\bibnamefont {Painter}},\ }\href
  {http://scitation.aip.org/content/aip/journal/apl/101/8/10.1063/1.4747726}
  {\bibfield  {journal} {\bibinfo  {journal} {Applied Physics Letters}\
  }\textbf {\bibinfo {volume} {101}},\ \bibinfo {eid} {081115} (\bibinfo {year}
  {2012})}\BibitemShut {NoStop}%
\bibitem [{\citenamefont {Datta}(1997)}]{Datta1997}%
  \BibitemOpen
  \bibfield  {author} {\bibinfo {author} {\bibfnamefont {S.}~\bibnamefont
  {Datta}},\ }\href@noop {} {\emph {\bibinfo {title} {Electronic Transport in
  Mesoscopic Systems}}}\ (\bibinfo  {publisher} {Cambridge University Press},\
  \bibinfo {year} {1997})\BibitemShut {NoStop}%
\bibitem [{\citenamefont {Weisskopf}\ and\ \citenamefont
  {Wigner}(1930)}]{Weisskopf1930}%
  \BibitemOpen
  \bibfield  {author} {\bibinfo {author} {\bibfnamefont {V.}~\bibnamefont
  {Weisskopf}}\ and\ \bibinfo {author} {\bibfnamefont {E.}~\bibnamefont
  {Wigner}},\ }\href {\doibase 10.1007/BF01336768} {\bibfield  {journal}
  {\bibinfo  {journal} {Zeitschrift f{\"u}r Physik}\ }\textbf {\bibinfo
  {volume} {63}},\ \bibinfo {pages} {54} (\bibinfo {year} {1930})}\BibitemShut
  {NoStop}%
\bibitem [{\citenamefont {Scully}\ and\ \citenamefont
  {Zubairy}(1997)}]{Scully1997}%
  \BibitemOpen
  \bibfield  {author} {\bibinfo {author} {\bibfnamefont {M.~O.}\ \bibnamefont
  {Scully}}\ and\ \bibinfo {author} {\bibfnamefont {M.~S.}\ \bibnamefont
  {Zubairy}},\ }\href
  {http://www.cambridge.org/gb/knowledge/isbn/item1143161/?site_locale=en_GB}
  {\emph {\bibinfo {title} {Quantum Optics}}}\ (\bibinfo  {publisher}
  {Cambridge University Press},\ \bibinfo {year} {1997})\BibitemShut {NoStop}%
\bibitem [{\citenamefont {Weiss}(2012)}]{Weiss2012}%
  \BibitemOpen
  \bibfield  {author} {\bibinfo {author} {\bibfnamefont {U.}~\bibnamefont
  {Weiss}},\ }\href {http://www.worldscientific.com/worldscibooks/10.1142/8334}
  {\emph {\bibinfo {title} {Quantum Dissipative Systems}}},\ \bibinfo {edition}
  {4th}\ ed.\ (\bibinfo  {publisher} {Word Scientific},\ \bibinfo {year}
  {2012})\BibitemShut {NoStop}%
\end{thebibliography}%
\end{document}